\documentclass{icrc2009}

\usepackage{graphicx}   
\usepackage{url}

\newcommand{\shorttitle}[1]%
{\markboth{Proceedings of the 31\MakeLowercase{$^{st}$} ICRC, {\L}\'{o}d\'{z} 2009}{#1} }
\newcommand{\etal}{\MakeLowercase{\textit{et al. }}} 

\begin{document}
\title{The Knee in the Cosmic Ray Energy Spectrum}

\author{\IEEEauthorblockN{Anatoly Erlykin\IEEEauthorrefmark{1} and
			  Arnold Wolfendale\IEEEauthorrefmark{2}}
                            \\
\IEEEauthorblockA{\IEEEauthorrefmark{1}P.N.Lebedev Physical Institute, Moscow, Russia}
\IEEEauthorblockA{\IEEEauthorrefmark{2}Department of Physics, Durham University, Durham, UK}}

\shorttitle{Erlykin \etal The Knee in CR}
\maketitle

\begin{abstract}
An update of the status of the knee in the cosmic ray energy spectrum at 3-4 PeV is 
presented. We argue that the evidence in favour of the presence of a 'single source' 
is even stronger than before.
\end{abstract}

 
\section{Introduction}
In 1997 we put forward the idea that the remarkable sharpness of the knee at 3-4 PeV in
 the cosmic ray (CR) energy spectrum is due to the dominant contribution of a single 
source. This sharpness could be the consequence of the sharp cutoff of the maximum 
accelerated energy in the source. Another argument in favour of the single source model
was the fine structure of the spectrum in the vicinity of the knee ie, not just a 
single smooth transition, albeit a sharp one. Specifically, we found evidence for two 
'knees' \cite{Erl1}. Initially we thought that the dominant primary nucleus in the knee
 was oxygen. In this case the second structure, observed at an energy about 3.5-4 times
 higher than the energy of the first knee (~usually just referred to as 'the knee'~), 
is due to primary iron. 4 years later we updated the analysis using the 40 
size spectra of extensive air showers (EAS) available at that time. We confirmed 
the conclusion that the observed sharpness of the knee is higher than expected in the 
classic diffusion model. We extended the analysis on the spectra of Cherenkov light 
from EAS and found the same fine structure as observed with the detectors of EAS 
charged particles \cite{Erl2}. Later on, we came to the conclusion that the most likely 
nucleus, dominant in the knee, was helium, not oxygen. In this case the second observed
structure in the energy spectrum at 12-16 PeV is not due to iron, but to oxygen 
\cite{Erl3}. If this is true, at even higher energies, about 40-50 PeV, one can expect 
another structure due to iron. \\
    The application of the theoretical model of the supernova explosion to the 
observations yielded an estimate of the most likely distance (230-350 pc) and age 
(84-350 kyears) of the source \cite{Erl4}. American astronomers found that the 
pulsar B0656+14, located inside the Monogem Ring supernova remnant (SNR), fits well 
this distance and age. Therefore this complex can be the candidate for the single 
source. Unfortunately, it is difficult to observe this source by means of the 
anisotropy of charged CR, since the particles with PV-rigidity have a giroradius less 
than 1 pc, which is two orders of magnitude less than the estimated distance to the 
source. Similarly, it is difficult to confirm the location by detecting the Monogem 
Ring with TeV gamma-rays since it is not a discrete, but a very extended, SNR with an 
angular diameter about 18$^\circ$. \\ 
The aim of this paper is to review the present situation around the knee and update the
 status of the single source model. \\
\section{New data}
\begin{figure}[!h]
\centering
\includegraphics[height=3.4in,angle=-90]{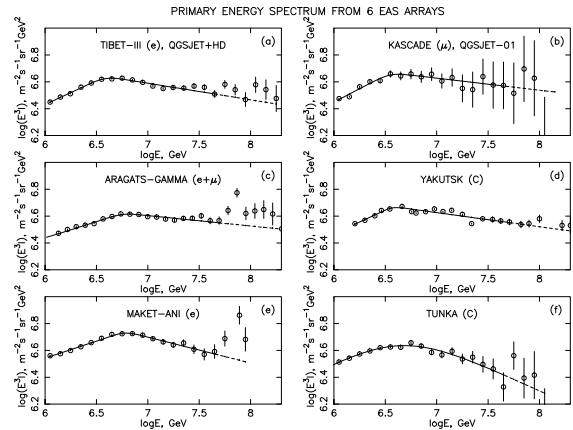}
\caption{\footnotesize Energy spectra of primary CR, measured by Tibet-III (a), KASCADE
 (b), GAMMA (c), Yakutsk (d), Maket-Ani (e) and Tunka (f) arrays. Symbols {\em e}, 
$\mu$ and {\em C} in brackets indicate the measured EAS component: electromagnetic, 
muon or Cherenkov light respectively. Notations QGSJET+HD ({\em heavy dominant}) in (a)
 and QGSJET-01 in (b) indicate the interaction model used for the transition from the 
measured parameters to the primary energy. Full lines are fits by the expression (2) 
with best fit parameters shown in the last five columns of Table 1. Dashed lines are 
extrapolations of these fits to the energy above the fitted range.}
\label{fig1}
\end{figure}
Since the beginning of this decade several new measurements of the CR energy spectrum 
have been published \cite{Tibet,Kasc,Gamma,Yakut,Maket,Tunka}. They are shown in Figure
 1. We do not present here the EAS-TOP+MACRO measurements \cite{EASTOP} since their 
authors showed spectra of light and heavy elements separately. We omitted also the
 results of KASCADE-Grande \cite{Kasc-Gr} since they are still preliminary.  
\subsection{Sharpness}
The position of the knee (~$logE^k_1$~) has been determined as the point with the 
maximum sharpness of the spectrum, the sharpness $S_1$ being defined as \cite{Erl1}
 \begin{equation}
S_1=-\frac{d^2(logI)}{d(logE)^2}
\end{equation}
The results are shown in columns 2 and 3 of Table 1.  \\
It is interesting to note that as a rule measurements of muons and Cherenkov light give
 higher sharpness than the electromagnetic component (~except for Tunka~). It 
is understandable since both muons and Cherenkov light are proxies of the energy lost 
by the cascade in the atmosphere which is quite close to the primary energy.  
\begin{table*}[!th] 
\label{table1}
\centering
\begin{tabular}{||c||c|c||c|c|c|c|c||} \hline
Array & $logE^k_1$ & $S_1$ & $logE^k_2$ & $\gamma$ & $\Delta \gamma$ & $\delta$ & $S_2$  \\ 
\hline
Tibet-III  & 6.60 & 1.34$\pm$0.21 & 6.59 & 2.64 & 0.48 & 8.85 & 2.44$\pm$0.77 \\
KASCADE    & 6.60 & 3.07$\pm$0.77 & 6.55 & 2.60 & 0.49 & 8.43 & 2.34$\pm$4.86 \\
GAMMA      & 6.61 & 1.19$\pm$0.30 & 6.76 & 2.76 & 0.32 & 10.0 & 1.84$\pm$0.43 \\
Yakutsk    & 6.69 & 3.56$\pm$0.65 & 6.57 & 2.64 & 0.46 & 15.7 & 4.17$\pm$4.31 \\
Maket-ANI  & 6.70 & 1.53$\pm$0.53 & 6.78 & 2.75 & 0.44 & 12.9 & 3.30$\pm$3.30 \\
Tunka      & 6.50 & 1.44$\pm$0.60 & 6.64 & 2.59 & 0.75 & 1.64 & 0.71$\pm$0.13 \\ \hline
\end{tabular}
\caption{\footnotesize Best fit parameters of the energy spectra presented in 
Fig.1. Columns 2 and 3 show parameters determined by expression (1), columns 4-8 - 
by expressions (2) and (3)}
\end{table*}

The result is that the primary energy spectrum is indeed rather sharp. In any case,
 Table 1 
shows that in all new measurements the sharpness of the knee is substantially 
higher than 0.3 which is the characteristic value for the Galactic Diffusion 
Model. \\ 
At this stage it is important to make some more remarks about what might be expected 
for the value of $S$ in the conventional case of many sources contributing. As pointed 
out in \cite{Erl5}, the value $S=0.3$ is, in fact, an upper limit for a 'normal 
composition'. Further analysis, involving the most likely generation of CR particles 
with a variety of exponents for their energy spectra - conditioned by different SNR 
expanding in regions of different density and magnetic field (~some of which will be 
increased by the CR shocks themselves~) - leads to $S$ value even lower than 0.3. The 
case for the Single Source Model, in which the sharpness is due to the vicinity of the 
single source ( SNR or pulsar ) with a sharp cutoff of the emission spectrum at the 
maximum acceleration energy is thus strengthened.
\subsection{Fine structure}
We define 'fine structure' of the spectrum as the existence of reliable deviations from
 the power law fits both below and above the knee. To find these deviations we fitted 
the spectra in Fig.1 by the expression proposed in \cite{Samvel}:
\begin{equation}
I(E)=AE^{-\gamma}(1+(\frac{E}{E^k})^\delta)^{-\frac{\Delta \gamma}{\delta}}
\end{equation}  
Here $\gamma$ is the power law index of the spectrum below the knee, which changed by
$\Delta \gamma$ above the knee. Between these regions there is a transition range 
which is described by the sharpness parameter $\delta$. The true sharpness $S_2$ is 
connected with $\delta$ as 
\begin{equation}
S_2 = \delta \Delta \gamma \frac{ln10}{4} 
\end{equation}
We have found best fit values for these parameters using the least squares method with 
MINUIT code \cite{MINUIT}. They are shown in the 5 last columns of Table 1. The energy
 range, in which we fitted spectra, and the fit itself, are shown by the full line in 
Fig.1. The linear extrapolation of the fit to higher energies is indicated by the 
dashed line. \\
Deviations of the actual intensities from this fit and its extrapolation for all 6 
spectra are shown in Fig.2a. In order to remove the difference between energy scale of 
all spectra and 
reveal the peculiarities in their {\em shape} we referred the deviations from the fit
(2) in the individual spectra to the individual energy of the knee $logE^k_2$. Mean 
values of the deviation are shown in Fig.2b. \\      
\begin{figure}[thb]
\centering
\includegraphics[height=3.4in,angle=-90]{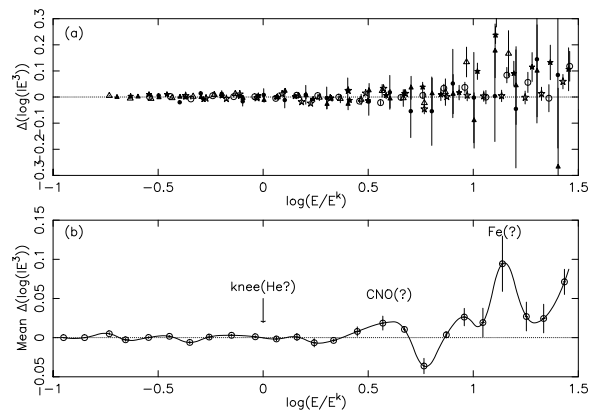}
\caption{\footnotesize Fine structure of the PCR energy spectrum. The irregularity at
the position of the knee, $log(E/E^k) = 0$, is not seen since the expression (2) gives 
a good fit of the spectrum in the knee region.} 
\label{fig:fig2}
\end{figure}
The irregularity at $log(E/E^k)=0.5-0.6$ found in \cite{Erl1,Erl2} is confirmed in 
the new spectra too. The progress in these new  measurements lets us proceed to 
higher energies. Here, a new feature can be seen at $log(E/E^k)=1-1.2$. It was  
first noticed in \cite{Gamma} and now confirmed by 5 other spectra. \\ 
If indeed $He$ nuclei dominate in the knee then the first irregularity at 
$log(E/E^k)=0.5-0.6$ can be referred to the $CNO$-group of nuclei and the second one at
$log(E/E^k)=1-1.2$ - to the $Fe$-group. If true, the existence of these groups is in 
favour of the single source model. The peaks marked {\em 'CNO(?)'} and {\em 'Fe(?)'} 
are at the correct places for the nuclei which are thought to form the bulk of the CR 
after hydrogen and helium. 
\subsection{Other evidence}
During the last year the PAMELA and ATIC collaborations claimed that they observed an 
excess of positrons \cite{Adrian1} and electrons ($e^-+e^+$) \cite{Chang} in the 
primary CR. The evidence was in the form of a sudden upturn in the positron spectrum at
 $\sim3-5$ GeV leading to a bump in the $e^-+e^+$ spectrum at $\sim500$ GeV. The peak 
is some 3-4 times the 'background level'formed by a smooth steepening of the spectrum 
from energies below the bump. The other measurements although showing somewhat smaller 
intensity in the bump confirmed an irregular behavior of the spectrum in this region
\cite{Abdo} and its sharp steepening at TeV energies above the bump \cite{HESS1,HESS2},
which creates the feature similar or even sharper than the knee at PeV energies. \\
These publications caused great interest and inspired many attempts at 
their explanation (~see references in \cite{Barger}~). The bulk of the 
proposed models suggested mechanisms in which the additional  
electrons and positrons were created by the interaction, annihilation or decay of dark 
matter particles. Other models proposed astrophysical scenarios with extra electrons 
and positrons created, accelerated and emitted by various astrophysical sources. We 
consider these latter scenarios as more likely not only because dark matter particles 
are still elusive, but such models face the difficulty of an absence of extra
 antiprotons in the PAMELA data \cite{Adrian2}, which should inevitably be produced in 
processes 
including dark matter particles. To suppress the production of antiprotons the models 
with dark matter include additional assumptions which make them more complicated and 
speculative.
\section{Discussion of the knee and of the 'electron bump'}
    We discuss astrophysical models here because they seem to us more realistic and 
according to our view they give support to our single source model. Here we present
 arguments in favor of such a view.
\begin{itemize}
\item The essence of the single source model is a concept that CR sources are 
non-uniformly distributed in space and time. As a consequence the CR energy spectrum 
observed at the Earth can carry traces of this non-uniformity, i.e. 
irregularities of some kind or a fine structure. In particular the substantial 
contribution of just the nearby and recent single source (~SNR or pulsar~) to the flux 
of CR protons and nuclei can be the cause of the knee at PeV energies.
\item The observed sharpness of the knee is due to several reasons: (i) the source is 
relatively close to the solar system and the energy spectrum of its CR is not distorted
by the propagation effects. Its shape is close to the slope of the production spectrum.
 Below the knee it is rather flat (~$\gamma \approx 2.1$~) compared with the bulk of CR
(~$\gamma \approx 2.7$~) and its contribution is more pronounced at high energies.   
(ii) The CR energy spectrum has a sharp cutoff at the maximum acceleration energy. 
(iii) If the {\em He} component is dominant at the knee it gives an additional 
sharpness since we must expect a gap between the {\em He} and {\em CNO} group of 
nuclei. (iv) Since the source is 'single' the smoothing effect on the knee sharpness 
due to the spread of characteristics inevitable in the case of multiple sources is at 
a minimum.  
\item The electron component of CR is even more sensitive to the presence of nearby and
 recent sources than protons and nuclei, since electrons from remote and old sources 
suffer not only from diffusive losses, but also from rising energy losses.  
Indeed, in our paper devoted to SNR and the electron component \cite{Erl6} we have 
shown how different electron spectra could be for different samples of the time-space 
 distribution of SNR in our Galaxy. We show this collection of electron spectra in 
Figure 3. The simulations were made assuming that the electrons are accelerated in
 the SNR in a similar manner to that for protons. It is seen that the fine structure of
 the spectrum appears already at energies above 100 Gev and clearly seen as bumps in 
the TeV region, which is due to the presence of recent and nearby SNR. 
\begin{figure}[thb]
\centering
\includegraphics[height=3.4in,angle=-90]{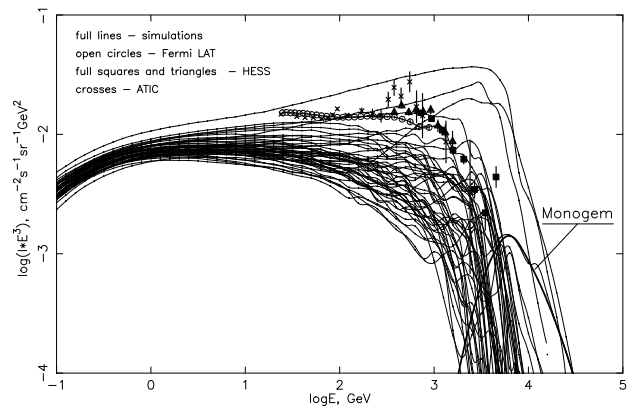}
\caption{\footnotesize Examples of predicted electron spectra for different time-space 
distributions of SNR. Experimental data: crosses - ATIC2 \cite{Chang}, open circles -
Fermi LAT \cite{Abdo}, full squares and triangles - HESS \cite{HESS1,HESS2}. 
No normalization has been applied. The difference by the factor of 2-3 in absolute 
intensities between most simulated samples and the experimental data is not important 
and can be easily reduced by reasonable adjusting the model parameters. The bumps in 
the simulated samples are generally at an energy higher than the observed
electron-positron bump, which is at $\sim$500 GeV. This is interpreted by us as 
indicating that these electrons+positrons are secondaries to 'our' SNR-accelerated 
protons and nuclei. The thick solid line in the TeV region shows the contribution to 
electrons expected from the Monogem Ring SNR \cite{Erl6}.} 
\label{fig:fig3}
\end{figure}
\item  In the absence of likely sources of positrons within the SNR, the observations 
dictate that the positrons come from another process. SNR and pulsars are usually 
immersed in an the envelopes of gas: 
remnants of the SN explosion (~SNR~), and pulsar wind nebulae (~PWN~). The accelerated 
particles interact with this gas and create electron-positron pairs. These secondary
positrons and electrons will have energies less than electrons directly accelerated and
 their ensuing bump will be at lower energy: tens of GeV is not unreasonable.
 \item the energy spectrum of electrons from nearby and recent sources
 would be as flat as the production spectrum of the accelerated particles and create a 
feature similar to the knee in the spectrum of primary CR nuclei. The sharpness of this
 'electron knee' in this scenario is supported by HESS measurements \cite{HESS1,HESS2} 
and is due to the sharpness of the knee for primary nucleons, rising energy losses of 
electrons and positrons at TeV energies and a small pile up of electrons which 
initially had an energy higher than the knee, but lost it during the propagation. 
The calculations made in some works support such a possibility 
\cite{Barger,Profumo,Hong,Shaviv,Malysh,Blasi,Fujita,Piran}. There are, however, 
attempts to give a methodical explanation of the PAMELA and ATIC excess by the rising 
contamination of positrons and electrons from primary protons \cite{Fazely,Schub}.
\item the magnitude of the bump from the single source with respect to the background 
in electrons can be greater than that in nuclei and this can be understood. In view of 
the rapidly increasing energy losses for electrons in the general interstellar medium, 
compared with only a diffusive loss for nuclei, the electron background is relatively 
lower. We estimate that the bump (ATIC) contains an energy of $\sim 10^{-5}$eVcm$^{-3}$
; for reference, our single source contributes about $2\cdot 10^{-4}$eVcm$^{-3}$ in the
 CR spectrum.
\end{itemize}
\section{Conclusions}
We have analysed the six new energy spectra which have appeared in the papers published
 since the last update of our single source model. All the previous findings 
(~sharpness and the fine structure of the knee~) are confirmed by this new data. The 
advance to a higher 
energy of about 10$^8$GeV lead us to confirm the existence of a new feature - another 
irregularity in the spectrum at energies of 50-80 PeV, claimed first in \cite{Gamma}.
 If the dominant contribution to the knee is due to primary {\em He}-nuclei, this new 
irregularity is just where primary iron nuclei should appear. \\
We consider that the latest findings of the irregularities in the electron and 
positron spectra (~'electron bump or knee'~) can have the same origin as the 'hadron 
knee' in the spectra of the primary nuclei, i.e. they are due to the non-uniformity of
the time-space distribution of CR sources, and if true, it is an additional support of 
the single source model. \\
\newpage
It is interesting to postulate that the last HESS point at $\sim$4500 GeV could 
represent part of a signature of SNR-accelerated electrons from a source such as 
Monogem Ring; certainly it is in the energy region where we expect a bump. \\
{\em Acknowledgments} \\
The authors are thankful to the Ralph Kohn Foundation for their interest and 
financial support of this work. \\

\end{document}